\documentclass[aps,twocolumn,showpacs,preprintnumbers,floats]{revtex4}\def\@cite#1#2{\textsuperscript{[{#1\if@tempswa , #2\fi}]}}

\usepackage{graphicx}
\usepackage{amsmath}
\usepackage{amsfonts}
\usepackage{amssymb}
\usepackage{color}
\usepackage{subfigure}
\usepackage{epsfig}
\usepackage{morefloats}
\usepackage{multirow}
\usepackage{graphicx,booktabs}
\usepackage{mathrsfs}
\usepackage{txfonts}
\usepackage{indentfirst}
\usepackage{graphicx,booktabs}

\usepackage{longtable,lscape}
\usepackage[colorlinks, citecolor=blue,anchorcolor=red,menucolor=red, linkcolor=red,filecolor=red,urlcolor=blue,frenchlinks=red]{hyperref}

\newcommand{\vsig}{\mbox{\boldmath$\sigma$\unboldmath}}

\begin{document}

\title{Further understanding the nature of $\Omega(2012)$ within a chiral quark model}

\author{Hui-Hua Zhong, Ru-Hui Ni, Mu-Yang Chen~\footnote {E-mail: muyang@hunnu.edu.cn},
Xian-Hui Zhong~\footnote {E-mail: zhongxh@hunnu.edu.cn}}

\affiliation{ Department of Physics, Hunan Normal University, and Key Laboratory of Low-Dimensional Quantum Structures and Quantum Control of Ministry of Education, Changsha 410081, China }

\affiliation{ Synergetic Innovation Center for Quantum Effects and Applications (SICQEA),
Hunan Normal University, Changsha 410081, China}

\author{  Ju-Jun Xie \footnote {E-mail: xiejujun@impcas.ac.cn} }
\affiliation{ Institute of Modern Physics, Chinese Academy of
Sciences, Lanzhou 730000, China} \affiliation{School of Nuclear
Science and Technology, University of Chinese Academy of Sciences,
Beijing 101408, China} \affiliation{School of Physics and
Microelectronics, Zhengzhou University, Zhengzhou, Henan 450001,
China}

%


\begin{abstract}

In our previous works, we have analyzed the two-body strong decays of
the low-lying $\Omega$ baryon states within a chiral quark model. The results show that the
$\Omega(2012)$ resonance favors the three-quark state with $J^P=3/2^-$ classified
in the quark model. With this assignment, in the present work we further study the three-body strong decay
$\Omega(2012)\to \Xi^*(1530)\bar{K} \to \Xi\pi\bar{K}$
and coupled-channel effects on $\Omega(2012)$ from nearby channels $\Xi \bar{K}$, $\Omega\eta$ and $\Xi^*(1530)\bar{K}$ within
the chiral quark model as well. It is found that the $\Omega(2012)$ resonance has a sizeable decay rate into the three-body final state
$\Xi\pi\bar{K}$. The predicted ratio $R_{\Xi\bar{K}}^{\Xi\pi\bar{K}}=\mathcal{B}[\Omega(2012)\to \Xi^*(1530)\bar{K}\to \Xi\pi\bar{K}]/\mathcal{B}[\Omega(2012)\to \Xi\bar{K}]\simeq 12\%$ is close to the up limit $11\%$ measured by the
Belle Collaboration in 2019, however, our predicted ratio is too small to be comparable with the recent data $0.97\pm 0.31$.
Furthermore, our results show that the coupled-channel effects on the $\Omega(2012)$ is not large, its components
should be dominated by the bare three-quark state, while the proportion of
the molecular components is only $\sim 16\%$. To clarify the nature of $\Omega(2012)$,
the ratio $R_{\Xi\bar{K}}^{\Xi\pi\bar{K}}$ is expected to be tested by other experiments.

\end{abstract}


\maketitle

\section{Introduction}

In 2018, Belle Collaboration observed a new
excited hyperon $\Omega(2012)^-$ decaying into $\Xi^0K^-$ and $\Xi^-\bar{K}^0$ with a mass of
$2012.4\pm 0.7(\mathrm{stat})\pm 0.6(\mathrm{syst})$ MeV and a width of $\Gamma=6.4^{+2.5}_{-2.0}(\mathrm{stat})\pm1.6(\mathrm{syst})$ MeV~\cite{Belle:2018mqs}. In 2021, evidence of $\Omega(2012)^-$ was observed in the $\Omega_c$ weak decay process
$\Omega_c\to \pi^+\Omega(2012)^-\to \pi^+(\Xi^0 K^-)$ at Belle~\cite{Belle:2021gtf} following the proposal in Ref.~\cite{Zeng:2020och}.
According to the previous mass spectrum predictions in various models and methods, such as the Skyrme model~\cite{Oh:2007cr}, quark model~\cite{Capstick:1986bm,Faustov:2015eba,Loring:2001ky,Liu:2007yi,Chao:1980em,Chen:2009de,An:2013zoa,Kalman:1982ut,Pervin:2007wa,An:2014lga},
lattice gauge theory~\cite{Engel:2013ig,Liang:2015bxr} and so on~\cite{Carlson:2000zr,Goity:2003ab,Schat:2001xr,Matagne:2006zf,Bijker:2000gq,Aliev:2016jnp},
the newly observed $\Omega(2012)$ may be a good candidate of the first orbital ($1P$) excitations
of $\Omega(1672)$.

The discovery of $\Omega(2012)$ immediately attracted a great deal of
attention from the hadron physics community in the literature.
Our group analyzed the Okubo-Zweig-Iizuka (OZI)-allowed two body strong decays of the low-lying $P$- and $D$-wave $\Omega$
baryon states within the chiral quark model~\cite{Liu:2019wdr,Xiao:2018pwe} and $^3P_0$ model~\cite{Wang:2018hmi}
by combining the mass spectrum analysis, and found that the $\Omega(2012)^-$ resonance favors the assignment of the $1P$-wave state with $J^P=3/2^-$.
Furthermore, the newly measured ratio $\mathcal{B}[\Omega_c\to \Omega(2012)\pi^+ \to
(\Xi\bar{K})^-\pi^+ ]/\mathcal{B}[\Omega_c\to \Omega \pi^+]$ at Belle can be well
understood as well within the three quark picture in a recent work~\cite{Wang:2022zja}.
Our conclusion is consistent with that based on the framework of QCD sum rules~\cite{Aliev:2018yjo,Aliev:2018syi}, the constituent quark model by including relativistic corrections~\cite{Arifi:2022ntc}, and the flavour SU(3) analysis~\cite{Polyakov:2018mow}.
However, in the literature the $\Omega(2012)$ was interpreted as a hadronic molecule~\cite{Hu:2022pae,Ikeno:2022jpe,Lin:2019tex,Lu:2020ste,Ikeno:2020vqv,Xie:2021dwe,Valderrama:2018bmv,Pavao:2018xub,Huang:2018wth,Gutsche:2019eoh}.
In the hadronic molecular picture, the decay rate into the three-body final state $\Xi \pi \bar{K}$ is
predicted to be similar to that of the two-body final state $\Xi \bar{K}$. While in the three-quark picture,
the decay rates for OZI-allowed two-body final state $\Xi \bar{K}$ should be dominant, the decay rate of
$\Omega(2012)^-\to \Xi^*(1530)\bar{K} \to \Xi \pi \bar{K}$ should be suppressed by the
intermediate hadron state $\Xi^*(1530)$.

To test the decay properties of $\Omega(2012)$ predicted within
different pictures, the three-body channel $\Xi \pi \bar{K} $
through the decay process $\Omega(2012)^-\to \Xi^*(1530)\bar{K}\to \Xi\pi \bar{K}$ was
observed by the Belle Collaboration. The first observation was carried out in 2019~\cite{Belle:2019zco}.
The collaboration observed no significant $\Omega(2012)$ signals in the three-body final
state $\Xi \pi \bar{K} $, and set an upper limit at 90 confidence
level on the branching fraction ratio
$R_{\Xi\bar{K}}^{\Xi\pi\bar{K}}\equiv \mathcal{B}[\Omega(2012)^-\to \Xi^*(1530)\bar{K}\to \Xi\pi\bar{K}]/\mathcal{B}[\Omega(2012)^-\to \Xi\bar{K}]<11\%$, which is consistent with expectation in the three-quark picture.
However, in the recent search, they observed significant $\Omega(2012)$
signals in the three-body decay process $\Omega(2012)^-\to \Xi^*(1530)^0K^-\to \Xi^-\pi^+K^-$~\cite{Belle:2022mrg}.
A rather large ratio $R_{\Xi\bar{K}}^{\Xi\pi\bar{K}}=0.97\pm 0.24\pm 0.07$ was extracted from the observations,
which seems to be consistent with expectation in the hadronic molecular picture.

In the previous works~\cite{Liu:2019wdr,Xiao:2018pwe}, we did not study the three-body decay
process $\Omega(2012)^-\to \Xi^*(1530)\bar{K}\to \Xi\pi \bar{K}$. It is unclear how large the decay rate
of the $\Xi\pi \bar{K}$ mode within our three-quark picture. Furthermore, the $\Omega(2012)$ as a
conventional three-quark state, it may strongly couple to the $\Xi (1530)\bar{K}$ channel,
then the $\Omega(2012)$ may contain significant molecular components due to
coupled-channel effects. To uncover these puzzles and better understand the nature of
$\Omega(2012)$, in the three quark picture we further study the three-body decays
of $\Omega(2012)\to \Xi \pi \bar{K}$ and the coupled-channel effects from nearby channels $\Xi \bar{K}$,
$\Omega\eta$ and $\Xi^{*}(1530)\bar{K}$. For self-consistency, we adopt the same framework, i.e.,
the chiral quark model~\cite{Manohar:1983md}, as our previous works for
the $\Omega(2012)$ studies~\cite{Liu:2019wdr,Xiao:2018pwe}.
In this model an effective chiral Lagrangian is
introduced to account for the quark-meson coupling at
the baryon-meson interaction vertex. The light pseudoscalar
mesons, i.e., $\pi,K$, and $\eta$, are treated as Goldstone bosons.
Since the quark-meson coupling is invariant under the chiral
transformation, some of the low-energy properties of QCD
are retained~\cite{Li:1994cy,Li:1997gd,Zhao:2002id}.

This paper is organized as follows. In Sec.~\ref{decay}, a brief review of the chiral quark model is given,
then by using this model the two-body and three body strong decays of $\Omega(2012)$ are estimated.
In Sec.~\ref{Coupled-channel effects}, the coupled-channel effects on the $\Omega(2012)$ are evaluated by combining the simplest version of the coupled-channel model with the chiral quark model. Finally, we give a short discussion and summary in Sec.~\ref{Summary}.

\begin{table}
\caption{The baryon wave functions involved in our calculations.
The $\psi_{n_{\rho}l_{\rho}m_{\rho}}^{\rho}$ and $
\psi_{n_{\lambda}l_{\lambda}m_{\lambda}}^{\lambda}$ represent the radial wave functions of $\rho$-mode and $\lambda$-mode, respectively, while the $\chi_{SS_z}^{\sigma}$ and $\phi_{\Omega,\Xi^*,\Xi}^{\sigma}$ expressed as spin and flavor wave functions of the baryon system, respectively.
The details can be found in Ref.~\cite{Xiao:2013xi}. }
\label{baryon wave function}
\begin{tabular}{l}
\hline\hline
$\Omega^*\left|1^2P_{ 3/2^{-}}\right\rangle
~=\frac{1}{\sqrt{2}}\left\{
\psi_{000}^{\rho}
\psi_{01m_{\lambda}}^{\lambda}
\chi_{\frac{1}{2}S_z}^{\lambda}
\phi_{\Omega}^{s}
+\psi_{01m_{\rho}}^{\rho}
\psi_{000}^{\lambda}
\chi_{\frac{1}{2}S_z}^{\rho}
\phi_{\Omega}^{s} \right\}$    \\
$\Omega\left|1^4S_{3/2^{+}}\right\rangle
~~=\psi_{000}^{\rho}
\psi_{000}^{\lambda}
\chi_{\frac{3}{2}S_z}^{s}
\phi_{\Omega}^{s}$     \\
$\Xi^{*}(1530)\left|1^4S_{3/2^{+}}\right\rangle
~=\psi_{000}^{\rho}
\psi_{000}^{\lambda}
\chi_{\frac{3}{2}S_z}^{s}\phi_{\Xi^{*}}^{s}$\\
$\Xi\left|1^2S_{1/2^{+}}\right\rangle
~~~=\frac{1}{\sqrt{2}} \left\{ \psi_{000}^{\rho}
\psi_{000}^{\lambda}
\chi_{\frac{1}{2}\frac{1}{2}}^{\rho}
\phi_{\Xi}^{\rho}
+\psi_{000}^{\rho}
\psi_{000}^{\lambda}
\chi_{\frac{1}{2}\frac{1}{2}}^{\lambda}
\phi_{\Xi}^{\lambda} \right\}$ \\
\hline\hline
\end{tabular}
\end{table}

\begin{table}[htp]
\caption{Masses (MeV) of the hadrons adopted in this work.}\label{HadronMass}
\begin{tabular}{lll}
\hline\hline
Hadron~~~~~~~~~~~  & $J^{P}$~~~~~~~~~~~ & Mass~(MeV)~~~~~ \tabularnewline
\hline
$\pi^{0},\pi^{\pm}$~~~~~~~~~~~  & $0^{-}$~~~~~~~~~~~  & $135.0,139.6$~~ \\
$\bar{K}^{0},K^{\pm}$~~~~~~~~~~~  & $0^{-}$~~~~~~~~~~~  & $497.6,493.6$~~ \\
$\eta$~~~~~~~~~~~  & $0^{-}$~~~~~~~~~~~  & $547.9$~~ \\
$\Xi^{-},\Xi^{0}$~~~~~~~~~~~  & $1/2^{+}$~~~~~~~~~~~  & $1321.7,1314.9$~~ \\
$\Xi^{*-},\Xi^{*0}$~~~~~~~~~~~  & $3/2^{+}$~~~~~~~~~~~  & $1535.0,1531.8$~~ \\
$\Omega^{-}$~~~~~~~~~~~ & $3/2^{+}$~~~~~~~~~~~ & $1671.7$~~ \\
$\Omega(2012)$~~~~~~~~~~~  & $3/2^{-}$~~~~~~~~~~~  & $2012.4$~~ \tabularnewline
\hline\hline
\end{tabular}
\end{table}

\section{strong Decay}\label{decay}

\subsection{two-body decay}

In the chiral quark model~\cite{Manohar:1983md}, the low-energy quark-pseudoscalar-meson interactions in the SU(3) flavor basis are represented by the effective Lagrangian~\cite{Li:1994cy,Li:1997gd,Zhao:2002id}
\begin{eqnarray}\label{Lagrangians}
\mathbf{{\cal L}}&=&\sum_j
\frac{1}{f_m}\bar{\psi}_j\gamma^{j}_{\mu}\gamma^{j}_{5}\psi_j\partial^{\mu}\phi_m.\label{coup}
\end{eqnarray}
In the above effective Lagrangians, $\psi_j$ represents the $j$th quark field in the hadron, $\phi_m$ is the pseudoscalar meson field, $f_m$ is the pseudoscalar meson decay constant. To match the nonrelativistic wave functions of the initial and final hadron states, we adopt the nonrelativistic form of the Lagrangians in following form,
\begin{eqnarray}\label{non-relativistic-expansST}
\mathbf{{\cal H}}_{m}^{nr}=\frac{\delta\sqrt{(E_f+M_f)(E_i+M_i)}}{f_m}\sum_j\Big\{\frac{\omega_m}{E_f+M_f}\vsig_j\cdot
\textbf{P}_f \nonumber\\
+ \frac{\omega_m}{E_i+M_i}\vsig_j \cdot
\textbf{P}_i-\vsig_j \cdot \textbf{q} +\frac{\omega_m}{2\mu_q}\vsig_j\cdot
\textbf{p}'_j\Big\}I_j \phi_m,
\end{eqnarray}
where $\omega_m$ and $\textbf{q}_m$ are the
energy and three momentum of the final state pseudoscalar meson, respectively; $E_i$ and $M_{i}$ are the energy and
mass of the initial heavy hadron, respectively; $E_f$ and $M_f$ represent the energy and mass of the final state heavy hadron, respectively;
$\textbf{p}_j$ is the internal momentum operator of the $j$th quark in the baryon system rest frame; $\vsig_j$ is the spin operator for the $j$th quark of the baryon system; and $\mu_q$ is a reduced mass expressed as $1/\mu_q=1/m_j+1/m'_j$ with $m_j$ and $m'_j$ for the masses of the $j$th quark in the initial and final baryons, respectively. The plane wave part of the emitted light meson is $\varphi_m=e^{-i\textbf{q}\cdot\textbf{r}_j}$, and $I_j$ is the flavor operator defined for the transitions in the SU(3) flavor space
\cite{Li:1997gd,Zhong:2007gp,Zhong:2008kd}. $\delta$ as a global parameter accounts for the strength of the quark-meson couplings. Here, we take the same value as that determined in Refs.~\cite{Xiao:2013xi,Xiao:2018pwe,Liu:2019wdr}, i.e., $\delta =0.576$.

For a excited baryon state, within the chiral quark model its two-body OZI-allowed strong decay amplitudes can be described by
\begin{eqnarray}\label{Amplitude}
\mathcal{M}\left[\mathcal{B} \rightarrow \mathcal{B}^{\prime} \mathbb{M}\right]
=\left\langle\mathcal{B}^{\prime}\left|\mathbf{{\cal H}}_{m}^{n r}\right| \mathcal{B}\right\rangle,
\end{eqnarray}
where $\left| \mathcal{B}\right\rangle$ and $\left|\mathcal{B'}\right\rangle$ are the wave functions of the initial and final baryon states, respectively. With derived decay amplitudes from Eq.(\ref{Amplitude}), the partial decay width for the $\mathcal{B} \rightarrow \mathcal{B}^{\prime} \mathbb{M}$ process can be calculated with
\begin{eqnarray}\label{TwoBodyDecay}
\Gamma=\frac{1}{8 \pi} \frac{\left|\mathbf{q}_m\right|}{M_{i}^{2}}\frac{1}{2J_i+1}\sum_{J_{i z} J_{f z}}\left|\mathcal{M}_{J_{i z} J_{f z}}\right|^{2},
\end{eqnarray}
where $J_{i z}$ and $J_{f z}$ represent the third components of the total angular momenta of the initial and final baryons, respectively.

\begin{figure}
\centering \epsfxsize=6.0 cm \epsfxsize=6.0 cm \epsfbox{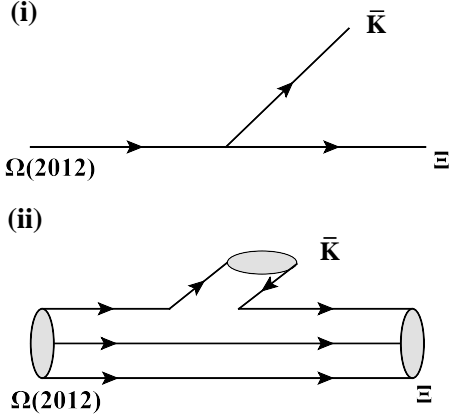} \vspace{0.4 cm} \caption{The OZI-allowed two-body decay process $\Omega(2012) \to \Xi K$ at (i) the hadronic level and (ii) the quark level.}\label{Omega2012twoBodyDecay}
\end{figure}

The OZI-allowed two-body strong decay for the $\Omega$ resonances
have been evaluated under the frame of the chiral quark model in our previous works~\cite{Liu:2019wdr,Xiao:2018pwe}. It is found that the newly observed $\Omega(2012)$ resonance favors the assignment of the $1P$-wave state with $J^P=3/2^-$ (i.e. $\Omega^*|1P_{3/2^-}\rangle$ listed in Table~\ref{baryon wave function}) in the $\Omega$ baryon family~\cite{Liu:2019wdr}. Considering the $\Omega(2012)$ as $\Omega^*|1P_{3/2^-}\rangle$, the $\Xi \bar{K}$ is the only OZI allowed two-body decay channel, which is shown in Fig.~\ref{Omega2012twoBodyDecay}, while the $\Omega(2012) \to \Xi^{*}(1530)\bar{K}$ decay process is forbidden since the mass of $\Omega(2012)$ lies below the $\Xi^{*}(1530)\bar{K}$ threshold. Thus, the total width
should be nearly saturated by the $\Xi \bar{K}$ channel.

By using the wave function calculated
from the potential model~\cite{Liu:2019wdr}, within the chiral quark model the partial width of $\Omega(2012)\to \Xi \bar{K}$
are predicted to be
\begin{eqnarray}\label{wdb}
\Gamma_{\Xi \bar{K}}=\Gamma_{\Xi^0 K^-}+\Gamma_{\Xi^- \bar{K}^0}\simeq(3.0+2.7) \ \ \mathrm{MeV},
\end{eqnarray}
which is consistent with the data $\Gamma_{exp}=6.4^{+2.5}_{-2.0}\pm 1.6$ MeV~\cite{ParticleDataGroup:2020ssz}.
It should be mentioned that in the calculations, to consist with the mass spectrum predictions in the potential model~\cite{Liu:2019wdr},
the constituent quark masses of $u/d$ and $s$ quarks are adopted to be $m_{u}=m_{d}=350$ MeV
and $m_{s}=600$ MeV. The meson decay constants for $\pi$, $K$ and $\eta$ are taken with $f_{\pi}=132$ MeV
and $f_{K}=f_{\eta}=160$ MeV. The masses of the final and initial states are taken the measured values from experiments~\cite{ParticleDataGroup:2020ssz}, which have been collected in Table~\ref{HadronMass}.
The spatial wave functions of the baryons appearing in the initial and final states are adopted the
harmonic oscillator form. For the $\Xi$ and $\Xi^*(1530)$ states, the harmonic oscillator
strength parameter $\alpha_{\rho}$ is taken as $\alpha_{\rho}=400$ MeV, while the parameter $\alpha_{\lambda}$
for the $\lambda$ oscillator is related to $\alpha_{\rho}$ with $\alpha_{\lambda}=\sqrt[4]{3m_u/(2m_s+m_u)}\alpha_{\rho}$.
For the $\Omega^*|1P_{3/2^-}\rangle$ state, we take $\alpha_{\rho}=\alpha_{\lambda}=411$ MeV,
which is fitted by reproducing the root-mean-square radius of the $\rho$-mode excitations from the potential model~\cite{Liu:2019wdr}.

\begin{figure}
\centering \epsfxsize=6.0 cm \epsfbox{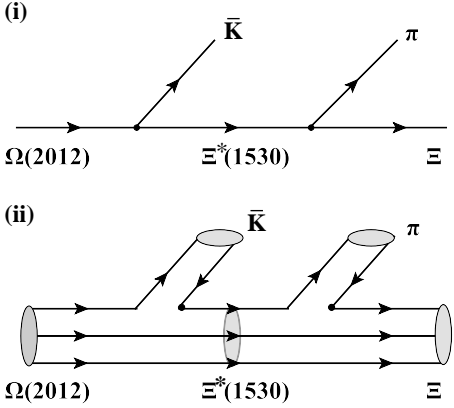} \vspace{0.4 cm} \caption{The cascade three-body decay process $\Omega(2012) \to \Xi^{*}(1530)K \to [\Xi \pi] K$ at (i) the hadronic level and (ii) the quark level.}\label{Omega2012ThreeBodyDecay}
\end{figure}

\begin{table}[htp]
\caption{The partial widths (MeV) of three-body final states for the $\Omega(2012)$ resonance. $\Gamma_i$ stands for the results
by considering the isospin breaking effects of the final states. While $\overline{\Gamma}_i$ stands for the results without the
isospin breaking effects, the mass of a hadron in the final states is adopted the average value of different charged states.}\label{Decay width}
\begin{tabular}{clcccccccccc}
\hline \hline
&Channel
&$\Xi^{-}\pi^{+}K^{-}$
&$\Xi^{0}\pi^{0}K^{-}$
&$\Xi^{-}\pi^{0}\bar{K}^{0}$
&$\Xi^{0}\pi^{-}\bar{K}^{0}$
&$\mathbf{Sum}$\\
&$\Gamma_i$ (MeV)
&$0.25$  &$0.18$    &$0.07$  &$0.17$ &$\mathbf{0.67}$ \\
&$\overline{\Gamma}_i$ (MeV)
&$0.21$  &$0.11$    &$0.11$  &$0.21$ &$\mathbf{0.64}$ \\
\hline\hline
\end{tabular}
\end{table}

\subsection{three-body decay}

Recently, the Belle Collaboration observed a rather
large three-body decay rate in the $\Omega(2012)^-\to \Xi^*(1530)^0K^-\to \Xi^-\pi^+K^-$ process.
This cascade decay process can be evaluated in the chiral quark model with the same parameter set as well.
For the cascade decay process $\Omega(2012) \to \Xi^{*}(1530)K \to [\Xi \pi] \bar{K}$ as shown in Fig.~\ref{Omega2012ThreeBodyDecay}, the decay amplitude can be expressed as
\begin{eqnarray}
\mathcal{M}_{\Omega(2012) \rightarrow \Xi \pi \bar{K}}=\frac{\mathcal{M}_{\Omega(2012) \to \Xi^{*}\bar{K}}\mathcal{M}_{\Xi^* \to \Xi\pi}}{2M_{\Xi^*}\left( M_{\Xi \pi}- M_{\Xi^*}+i \tilde{\Gamma}_{\Xi^* \to \Xi\pi} / 2\right)},
\end{eqnarray}
where $\mathcal{M}_{\Omega(2012) \to \Xi^{*}(1530)\bar{K}}$ and $\mathcal{M}_{\Xi^* \rightarrow \Xi\pi}$ are the amplitudes defined in Eq.(\ref{Amplitude}). It should be emphasized that the $\Xi^*(1530)$ baryon mass $M_{\Xi^*}$ in $\mathcal{M}_{\Omega(2012) \to \Xi^{*}K}$ and $\mathcal{M}_{\Xi^* \to \Xi\pi}$ is replaced with the invariant mass $M_{\Xi\pi}$ of $\Xi\pi$ system since the unstable baryon $\Xi^*(1530)$ can be slightly off-shell. Adopting the wave functions listed in Table~\ref{baryon wave function}, one can work out the $\mathcal{M}_{\Omega(2012) \to \Xi^{*}(1530)\bar{K}}$ and $\mathcal{M}_{\Xi^* \rightarrow \Xi\pi}$ within the chiral quark model.
Moreover, $\tilde{\Gamma}_{\Xi^*(1530) \rightarrow \Xi\pi}$ also depends on the invariant mass of $M_{\Xi\pi}$, and it can be obtained by Eq.(\ref{TwoBodyDecay}).

Then the three-body decay width of the $\Omega(2012)$ resonance in their rest frame was obtained as
\begin{eqnarray}\label{GammaMXiPi}
d \Gamma_{\Xi \pi \bar{K}}=\frac{M_{\Xi \pi}^{2}}{M_{\Omega^*}^{2}} \frac{|\mathbf{q}_{K}|}{4 \pi^{2}} \frac{\tilde{\Gamma}_{\Xi^{*} \rightarrow \Xi \pi}\overline{\left|\mathcal{M}_{\Omega(2012) \rightarrow \Xi^{*} \bar{K}}\right|^{2}}}{4 M_{\Xi^{*}}^{2}\left[\left(M_{\Xi \pi}-M_{\Xi^{*}}\right)^{2}+\tilde{\Gamma}_{ \Xi^{*} \rightarrow \Xi \pi}^{2}/4 \right]} d M_{\Xi \pi},
\end{eqnarray}
where the absolute value of the momentum for the final $K$ meson is give by
\begin{eqnarray}
|\mathbf{q}_{K}|=\frac{\sqrt{\left[M_{\Omega^*}^{2}-\left(m_{K}+M_{\pi \Xi}\right)^{2}\right]\left[M_{\Omega^*}^{2}-\left(m_{K}-M_{\pi \Xi}\right)^{2}\right]}}{2 M_{\Omega^*}}.
\end{eqnarray}
Integrating the Eq.(\ref{GammaMXiPi}) from  $M_{\Xi}+m_{\pi}$ to $M_{\Omega^*}-m_{K}$, one can easily work out the three-body decay width of the cascade decay process $\Omega(2012) \to \Xi^{*}(1530)\bar{K} \to [\Xi \pi] \bar{K}$. Our results have been listed
in Table~\ref{Decay width}.

The $\Omega(2012)$ as the $\Omega^*|1P_{3/2^-}\rangle$ state has a sizeable partial decay width
into the three-body final state $\Xi \pi \bar{K}$,
\begin{eqnarray}
\Gamma_{\Xi\pi\bar{K}}\simeq 0.67 \ \ \mathrm{MeV},
\end{eqnarray}
combined it with the two-body decay width predicted in Eq.(\ref{wdb}), the total width is predicted to be
\begin{eqnarray}
\Gamma_{total}\simeq 6.37 \ \ \mathrm{MeV},
\end{eqnarray}
which is in good agreement with the data $\Gamma_{exp}=6.4^{+2.5}_{-2.0}\pm 1.6$ MeV~\cite{ParticleDataGroup:2020ssz}.
The predicted ratio
\begin{eqnarray}
R_{\Xi\bar{K}}^{\Xi\pi\bar{K}}\equiv\frac{\mathcal{B}[\Omega(2012)\to \Xi^*(1530)\bar{K}\to \Xi\pi\bar{K}]}{\mathcal{B}[\Omega(2012)\to \Xi\bar{K}]}\simeq 12\%,
\end{eqnarray}
is slightly larger than the previous data $(6.0\pm 3.7\pm 1.3)\%$ measured by the Belle Collaboration in 2020~\cite{Belle:2019zco},
however, is a factor of $\sim4-8$ smaller than their recent data $0.97\pm 0.31$ with improved selection criteria~\cite{Belle:2022mrg}.
In the recent work~\cite{Arifi:2022ntc}, the three body decay of
$\Omega(2012)\to \Xi\pi\bar{K}$ was also studied within a constituent
quark model by including relativistic corrections, a small ratio $R_{\Xi\bar{K}}^{\Xi\pi\bar{K}}\simeq 4.5\%$
is obtained with the $\Omega^*|1P_{3/2^-}\rangle$ assignment. It should be mentioned that the isospin
breaking effects on the partial widths of the three-body decays are obvious. The results in Table~\ref{Decay width}
show that there are significant differences between the partial widths with isospin breaking effects and
that without isospin breaking effects.

The recent measured ratio $R_{\Xi\bar{K}}^{\Xi\pi\bar{K}}$ at Belle~\cite{Belle:2022mrg} seems to
be compatible with the molecular interpretation for the $\Omega(2012)$ proposed in Refs.~\cite{Valderrama:2018bmv,Pavao:2018xub,Huang:2018wth,Gutsche:2019eoh}
where similar branching fractions $\Omega(2012)$ decays into $\Xi \pi \bar{K}$ and $\Xi \bar{K}$
were predicted.

In the following, we will further explore the molecular components of
$\Omega(2012)$ due to coupled-channel effects.

\begin{figure}
\centering \epsfxsize=6.0 cm \epsfbox{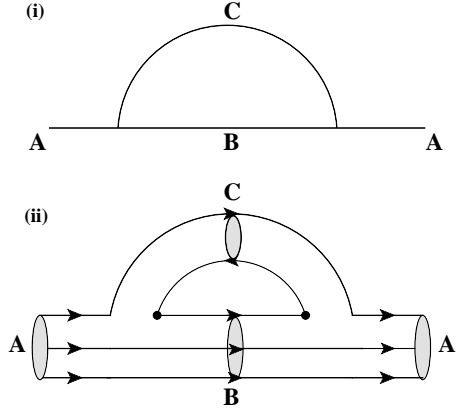} \vspace{0.4 cm} \caption{ The hadronic loop for a bare baryon state
$|A\rangle$ coupling to two-hadron continuum $BC$ at (i) the hadronic level and (ii) the quark level.}\label{Coupledchannel}
\end{figure}

\section{Coupled-channel effects}\label{Coupled-channel effects}

First, we give a brief review of the coupled-channel model adopted in this work. A bare baryon state $|A\rangle$ predicted in the quark model can couple to the two-hadron continuum $BC$ via hadronic loops, as shown in Fig.~\ref{Coupledchannel}. In the simplest version of the coupled-channel model~\cite{Morel:2002vk,Lu:2017hma,Kalashnikova:2005ui,Liu:2011yp,Lu:2016mbb}, the wave function of the physical state is given by
\begin{equation}
\begin{aligned}
| \Psi \rangle =c_A |A\rangle + \sum_{BC}\int c_{BC}(\mathbf{p})d^3\mathbf{p} |BC,\mathbf{p}\rangle ,
\end{aligned}
\end{equation}
where $\mathbf{p}=\mathbf{p}_B=-\mathbf{p}_C$ is final two-hadron relative momentum in the initial hadron static system, $c_A$ and $c_{BC}(\mathbf{p})$ denote the probability amplitudes of the bare valence state $|A\rangle$ and $|BC,\mathbf{p}\rangle$ continuum, respectively.

The coupling between the bare state $|A\rangle$ and the continuum components sectors $|BC,\mathbf{p}\rangle$ is achieved by creating light quark pairs. This coupling, as an effective coupling for the quark-meson interactions, can be described with
the chiral interactions given in Eq.(\ref{non-relativistic-expansST}) in the chiral quark model. Thus, the full Hamiltonian of the physical state $| \Psi \rangle$ can be written as
\begin{equation}
\begin{aligned}
\mathbf{{\cal H}} = \left( \begin{matrix}\mathbf{{\cal H}}_0~~~~~~\mathbf{{\cal H}}_m^{nr}
\\  \mathbf{{\cal H}}_m^{nr}~~~~~~\mathbf{{\cal H}}_c \end{matrix}\right),
\end{aligned}
\end{equation}
where $\mathbf{{\cal H}}_0$ is the Hamiltonian of the bare state $|A\rangle$ in the
potential model, while $\mathbf{{\cal H}}_c$ is the Hamiltonian for the continuum state $|BC,\mathbf{p}\rangle$.
Neglecting the interaction between the hadrons $B$ and $C$, one has
\begin{equation}
\begin{aligned}
\mathbf{{\cal H}}_{c}|BC,\mathbf{p}\rangle&=E_{BC}|BC,\mathbf{p}\rangle,
\end{aligned}
\end{equation}
where $E_{BC}=\sqrt{m_B^2+p^2}+\sqrt{m_C^2+p^2}$ represents the energy of $BC$ continuum.

The Schr\"{o}dinger equation of a mixed system can be written as
\begin{equation}\label{coupled-channel equation}
\begin{aligned}
&\left( \begin{matrix}\mathbf{{\cal H}}_0~~~~~~\mathbf{{\cal H}}_m^{nr}
\\  \mathbf{{\cal H}}_m^{nr}~~~~~~\mathbf{{\cal H}}_c \end{matrix}\right)
~\left( \begin{matrix} c_A |A\rangle
\\ \sum_{BC}\int c_{BC}(\mathbf{p}) d^3\mathbf{p} |BC,\mathbf{p}\rangle \end{matrix}\right)~~~~~~~~~~~~~~~~~~~~~~~~~~~~~~~\\
&~~~~~~~~~~~~~~~~~~~~~~~~~~~~~
=M \left( \begin{matrix} c_A |A\rangle
\\  \sum_{BC}\int c_{BC} (\mathbf{p}) d^3\mathbf{p}|BC,\mathbf{p}\rangle  \end{matrix}\right).
\end{aligned}
\end{equation}
From Eq.(~\ref{coupled-channel equation}), we have
\begin{eqnarray}\label{coupled-channel equation1}
\langle A | \mathbf{{\cal H}} | \Psi \rangle &=&c_A M \nonumber \\
    &=& c_A M_A+ \sum_{BC}\int c_{BC}(\mathbf{p})d^3\mathbf{p} \langle A | \mathbf{{\cal H}}_m^{nr} | BC,\mathbf{p}\rangle ,
\end{eqnarray}
\begin{eqnarray}\label{coupled-channel equation2}
\langle BC,\mathbf{p} | \mathbf{{\cal H}} | \Psi \rangle &=&c_{BC}(\mathbf{p}) M \nonumber\\
 &&=c_{BC}(\mathbf{p}) E_{BC}+ c_A\langle BC,\mathbf{p}| \mathbf{{\cal H}}_m^{nr} | A \rangle.
\end{eqnarray}
Deriving $c_{BC}(\mathbf{p})$ from Eq.(\ref{coupled-channel equation2}), and substituting it into Eq.(\ref{coupled-channel equation1}), we get a coupled-channel equation
\begin{equation}\label{M=MA+Delta M}
\begin{aligned}
M=M_A+\Delta M(M),
\end{aligned}
\end{equation}
where the mass shift $\Delta M(M)$ is given by
\begin{eqnarray}\label{ReDeltaM}
\Delta M(M) &&= \mathrm{Re} \sum_{BC}\int_0^{\infty} \frac{|\langle BC,\mathbf{p} |\mathbf{{\cal H}}_m^{nr}| A \rangle|^2}{(M-E_{BC})} p^2 dp d\Omega_p,\\
    &&= \mathrm{Re} \sum_{BC}\int_0^{\infty} \frac{\overline{|\mathcal{M}_{A\to BC}(\mathbf{p})|^2}}{(M-E_{BC})} p^2 dp ,
\end{eqnarray}
and $M_A$ is the bare mass of the baryon state $|A\rangle$ obtained from the potential model. From Eqs.(\ref{M=MA+Delta M}) and (\ref{ReDeltaM}), the physical mass $M$ and the bare state mass shift $\Delta M$ can be determined simultaneously. It should be mentioned that when we calculate the mass shift $\Delta M$ by using the Eq.(\ref{ReDeltaM}), the contribution in the higher $\mathbf{p}$ region may be nonphysical because the quark pair production rates via the non-perturbative interaction $\mathbf{{\cal H}}_m^{nr}$ should be strongly suppressed~\cite{Morel:2002vk}. In the chiral quark model, the chiral interaction $\mathbf{{\cal H}}_m^{nr}$ is only applicable to the low $\mathbf{p}$ region. To eliminate the nonphysical contributions, in our calculations, we cut off the momentum $p$ at the inflection point of $\Delta M(\mathbf{p})$ function as followed by our previous work~\cite{Ni:2021pce}. It should be noted that due to the various locations of the inflection point, the cut-off momentum for each channel varies.

\begin{table}[htp]
\caption{The proportions of different components and mass shift for the $\Omega(2012)$ resonance in the coupled-channel model.}\label{CoupledChannelResults}
\begin{tabular}{lccccccc}
\hline\hline
Channel & $\Xi \bar{K}$~~ & $\Xi^{*}(1530)\bar{K}$~~ & $\Omega\eta$~~ & \textbf{Baryon core}~~  &$\mathbf{Total}$~~\\
\hline
$\Delta M $ (MeV) & $-2.2$ & $-9.4$ & $-3.0$&$\mathbf{...}$  &$\mathbf{-14.6}$ \\
Proportion $(\%)$ & $1.9$ & $13.7$ & $0.8$&$\mathbf{83.6}$  &$\mathbf{100}$\\
\hline\hline
\end{tabular}
\end{table}

To estimate the component of baryon core and baryon-meson continuum components in the physical state, for below-threshold states, deriving $c_{BC}(\mathbf{p})$ from Eq.(\ref{coupled-channel equation2}), and substituting it into normalized constants containing two Fock components,
\begin{equation}\label{normalization condition}
\begin{aligned}
|c_A|^2+\sum_{BC}\int |c_{BC}(\mathbf{p})|^2 d^3\mathbf{p} =1,
\end{aligned}
\end{equation}
we get the probability of hadron $A$ component,
\begin{equation}\label{cA modular square(M<mB+mC)}
P_A=|c_A|^2= \left( 1+ \sum_{BC}  \int \frac{\overline{|\mathcal{M}_{A\to BC}(\mathbf{p})|^2}}{(M-E_{BC})^2} p^2 dp \right)^{-1}.
\end{equation}
Further, combining Eq.(\ref{coupled-channel equation2}) and Eq.(\ref{cA modular square(M<mB+mC)}), we can obtain the probability of the $|BC,\mathbf{p}\rangle$ continuum components,
\begin{equation}\label{cBC modular square(M<mB+mC)}
\begin{aligned}
P_{BC}=|c_A|^2 \int \frac{\overline{|\mathcal{M}_{A\to BC}(\mathbf{p})|^2} }{(M-E_{BC})^2} p^2 dp .
\end{aligned}
\end{equation}
The Eqs.(\ref{cA modular square(M<mB+mC)}) and (\ref{cBC modular square(M<mB+mC)}) can give the probability of the bare baryon components and $BC$ hadron pair components, in the coupled-channel effects. However, the integral in Eqs.(\ref{cA modular square(M<mB+mC)}) and (\ref{cBC modular square(M<mB+mC)}) is only well defined for $M<m_B+m_C$. For states above the $BC$ hadron pair threshold, as suggested by Ref.~\cite{Xie:2021dwe}, we replace the real $M$ with its complex counterpart, $M+i\Gamma/2$, for coupling channel calculation, where $\Gamma$ is the decay width of the physical state.

By using the above coupled-channel model, we study the coupled-channel effects on the
$\Omega(2012)$. The bare state $|A\rangle $ is considered to be the $1P$-wave state with $J^P=3/2^-$
(i.e. $\Omega^*|1P_{3/2^-}\rangle$) predicted within the quark model.
Three nearby channels $\Xi \bar{K}$, $\Omega\eta$ and $\Xi^{*}(1530)\bar{K}$ are considered.
The strong coupling amplitudes $\langle \Xi K/\Xi^{*}(1530)\bar{K}/\Omega\eta |\mathbf{{\cal H}}_m^{nr} |\Omega^*1P_{3/2^-}\rangle$
can be easily obtained within the chiral quark model by using the same parameter set.
With which, we further extract out the mass shift of the bare state $|\Omega^*1P_{3/2^-}\rangle$ due to the coupled-channel
effects and the proportion of various components.
Our results have been given in Table~\ref{CoupledChannelResults}.

It is found that, the coupled-channel effects on the $\Omega(2012)$ is not large.
There is a small mass shift of $\Delta\simeq -15$ MeV, which mainly contributed by
the $S$-wave channel $\Xi^{*}(1530)\bar{K}$. The components of $\Omega(2012)$ are dominated by
the bare quark model state $\Omega^*|1P_{3/2^-}\rangle$, its proportion can reach up
to about $84\%$; the proportion of the continuum components is only $\sim 16\%$.
In other words if the $\Omega(2012)$ resonance is a molecular dominant state, it cannot originate
from the three quark state $\Omega^*|1P_{3/2^-}\rangle$.


\section{Discussion and Summary}\label{Summary}

In our quark model study, with the $\Omega^*|1P_{3/2^-}\rangle$ assignment, both the
observed mass and width of $\Omega(2012)$ can be well explained~\cite{Liu:2019wdr,Wang:2018hmi,Xiao:2018pwe}.
Furthermore, the newly measured ratio $\mathcal{B}[\Omega_c\to \Omega(2012)\pi^+ \to
(\Xi\bar{K})^-\pi^+ ]/\mathcal{B}[\Omega_c\to \Omega \pi^+]$ at Belle can be well
understood as well in our recent work ~\cite{Wang:2022zja}.
The decay width is nearly saturated by two-body decay channel $\Xi K$. In the present work,
to further understand the nature of $\Omega(2012)$, in the three quark picture we study the three-body decays
of $\Omega(2012)\to \Xi \pi \bar{K}$ and coupled-channel effects from nearby channels $\Xi \bar{K}$,
$\Omega\eta$ and $\Xi^{*}(1530)\bar{K}$.

It is found that considering the $\Omega(2012)$ resonance as a conventional
$1P$-wave state $\Omega^*|1P_{3/2^-}\rangle$, there are sizeable decay rates into the
three-body final state $\Xi \pi \bar{K}$. The partial width and branching fraction is predicted
to be $\Gamma_{\Xi \pi \bar{K}}\simeq 0.67$ MeV and $\mathcal{B}[\Omega(2012)\to \Xi \pi \bar{K}]\simeq 11\%$.
The predicted ratio $R_{\Xi\bar{K}}^{\Xi\pi\bar{K}}=\mathcal{B}[\Omega(2012)^-\to \Xi^*(1530)\bar{K}\to \Xi\pi\bar{K}]/\mathcal{B}[\Omega(2012)^-\to \Xi\bar{K}]\simeq0.12$ is close to the previous data $(6.0\pm 3.7\pm 1.3)\%$ measured by the Belle Collaboration in 2019~\cite{Belle:2019zco}, however, is too small to be comparable with recent data $0.97\pm 0.31$~\cite{Belle:2022mrg}.

The newly measured ratio $R_{\Xi\bar{K}}^{\Xi\pi\bar{K}}=0.97\pm 0.31$ at Belle seems to
be consistent with the molecular interpretation for the $\Omega(2012)$ proposed in Refs.~\cite{Valderrama:2018bmv,Pavao:2018xub,Huang:2018wth,Gutsche:2019eoh}.
However, our results show that the coupled-channel effects on the $\Omega(2012)$ is not large,
if it originates from the $1P$-wave three-quark state $\Omega^*|1P_{3/2^-}\rangle$.
In this case, the components of $\Omega(2012)$ are dominated by
the bare quark model state $\Omega^*|1P_{3/2^-}\rangle$, its proportion can reach up
to about $84\%$.

The latest observations at Belle~\cite{Belle:2022mrg} make the situation
more complicated for understanding the nature of $\Omega(2012)$. The
ratio $R_{\Xi\bar{K}}^{\Xi\pi\bar{K}}$ is expected to be tested by other experiments.
Furthermore, the radiative decay of $\Omega(2012)\to \Omega \gamma$ is also expected to be
measured in future experiments. In the three quark picture, the branching fraction
is predicted to be $\mathcal{B}[\Omega(2012)\to \Omega \gamma]\simeq
1.7\times 10^{-3}$~\cite{Liu:2019wdr}, which can be used to test the nature of $\Omega(2012)$.

\section*{Acknowledgement}

This work is supported by the National Natural Science Foundation of China (Grants No.12175065, No. 12005060, No. 12075288). Ju-Jun Xie is also
supported by the Youth Innovation Promotion Association CAS.

\bibliographystyle{unsrt}

\end{document}